\documentclass[lettersize,journal]{IEEEtran}
\usepackage{amsmath,amsfonts}
\usepackage{algorithmic}
\usepackage{algorithm}
\usepackage{array}
\usepackage{textcomp}
\usepackage[table,xcdraw]{xcolor}
\usepackage{stfloats}
\usepackage{url}
\usepackage{siunitx}
\usepackage{verbatim}
\usepackage{graphicx}
\usepackage{subcaption}
\usepackage{cite}
\usepackage{comment}
\begin{document}

\title{Characterization of Event-Based Vision Sensors for High-Speed Optical Instrumentation}

\author{Tomás~Lopes,
        Joana~M.~Teixeira,
        Tiago~D.~Ferreira,
        Catarina~S.~Monteiro,
        Pedro~A.~S.~Jorge,
        and~Nuno~A.~Silva%
\thanks{Tomás Lopes, Joana M. Teixeira, Tiago D. Ferreira, Catarina S. Monteiro, Pedro A. S. Jorge, and Nuno A. Silva are with the Center for Applied Photonics, INESC TEC, Rua do Campo Alegre 687, 4169-007, Porto, Portugal}
\thanks{Tomás Lopes, Joana M. Teixeira, Pedro A. S. Jorge, and Nuno A. Silva are also with the Department of Physics and Astronomy, Faculty of Sciences, University of Porto, Rua do Campo Alegre s/n, 4169-007, Porto, Portugal.}%
\thanks{Tomás Lopes and Joana M. Teixeira contributed equally to this work.}}

\markboth{Journal of \LaTeX\ Class Files,~Vol.~14, No.~8, August~2021}%
{Shell \MakeLowercase{\textit{et al.}}: A Sample Article Using IEEEtran.cls for IEEE Journals}


\maketitle

\begin{abstract}
Event-based vision sensors provide asynchronous event generation and microsecond timestamp resolution, which may be useful for high-speed optical measurements. However, precise event timestamps do not necessarily guarantee accurate reconstruction of temporally varying optical signals, particularly under dense and spatially extended illumination, imposing operational limits when used as optical interrogators that remain underexplored in the literature. To address this knowledge gap, this work presents a systematic, quantitative characterization of the temporal response and waveform reconstruction fidelity of an IMX636-based event camera under both controlled sinusoidal and pulsed optical excitation. For this, frequency-domain measurements are first used to evaluate modulation response, event-rate behavior, polarity balance, and spectral reconstruction fidelity over a wide range of illumination conditions and region-of-interest geometries. Then, complementary pulse-based measurements quantify first-event latency, response duration, recovery dynamics, and pulse-width reconstruction accuracy under rapidly repeated excitation, showing that optical transitions can be detected with first-event latencies below \SI{5}{\micro\second}. However, the complete event response extends over significantly longer timescales due to photoreceptor dynamics, refractory behavior, and readout serialization. Under high-frequency modulation and short-pulse excitation, the reconstructed waveforms progressively degrade because of temporal spreading and imbalance between positive and negative event generation. The measurements further demonstrate that the temporal fidelity of the reconstructed signal depends strongly on the geometry and spatial activity of the selected region of interest. All in all, these results establish practical temporal-response limits for event-based optical measurements and provide relevant guidelines for effectively capitalizing on these devices as asynchronous optical measurement systems.

\end{abstract}


\section{Introduction}
Event-based vision sensors (EVS), also known as neuromorphic or dynamic vision sensors, implement an asynchronous sensing architecture in which individual pixels independently generate events whenever the temporal variation in logarithmic illumination exceeds a predefined contrast threshold \cite{lichtsteiner2008,gallego2020}. In contrast to conventional frame-based image sensors that sample the entire focal plane at fixed exposure intervals, EVS encode visual information as a sparse stream of timestamped events containing pixel coordinates, polarity, and temporal information. By combining logarithmic photoreceptor response with event-driven readout, these sensors achieve microsecond-scale temporal resolution, reduced data redundancy, and high dynamic range, enabling the measurement of rapidly varying optical signals \cite{gallego2020, lenero2018applications}.

Originally developed for dynamic scene analysis and low-latency visual sensing \cite{zafar2025applications,vasco2017independent,mueggler2017event, gallego2015event}, EVS have recently been applied to high-speed optical measurement systems, including optical fiber sensing, spectroscopy, microscopy, interferometry, and other high-bandwidth optical measurements where the relevant information is encoded in temporally varying optical signals rather than in conventional image formation \cite{ge2020dynamic,Su2025,Sopek2026}. For example, acoustic signal reconstruction has been demonstrated through the detection of speckle variations in multimode optical fibers and vibrating surfaces \cite{LOPES2026117985,cai2025event2audio}, while event-based spectroscopy has enabled microsecond-resolved spectral reconstruction at acquisition rates beyond the practical limits of conventional frame-based spectrometers \cite{teixeira2025}. In these applications, the relevant quantity is often the temporal evolution of the optical signal itself, requiring accurate reconstruction of modulation frequency, pulse timing, or waveform structure from the recorded event stream.

However, precise event timestamps do not necessarily imply accurate temporal reconstruction of optical waveforms. In many optical measurement systems, illumination changes occur across spatially extended regions of the sensor, producing dense, highly correlated event activity. Under these conditions, the measured response depends not only on the contrast-detection mechanism of individual pixels, but also on photoreceptor dynamics, refractory behavior, event thresholding, and readout serialization at board level. The response of event-based sensors is also known to depend strongly on illumination level, bias configuration, and spatial activity \cite{sengupta2024demystifying}. Consequently, the temporal fidelity of the reconstructed signal may differ substantially from the nominal timestamp resolution of the sensor. Despite the growing interest in neuromorphic optical sensing, the response of event-based cameras to dense, globally modulated optical excitation remains comparatively unexplored, particularly regarding how sensor architecture and readout mechanisms constrain the temporal information that can be reliably recovered from the event stream.

Aiming to tackle this knowledge gap and pave the way for future applications for event-based vision sensors, we present in this manuscript a systematic and quantitative characterization of the temporal response and waveform reconstruction fidelity of an IMX636-based event camera under controlled sinusoidal and pulsed optical excitation. Frequency-domain measurements are used to evaluate modulation response, event-rate behavior, polarity balance, and spectral reconstruction fidelity over a wide range of illumination conditions and region-of-interest geometries. Complementary pulse-based measurements are used to quantify first-event latency, response duration, recovery dynamics, and pulse-width reconstruction accuracy under rapidly repeated excitation. The results demonstrate that microsecond-scale event detection does not necessarily imply microsecond-scale waveform fidelity, and establish practical temporal-response limits and useful guidelines for event-based optical measurement systems.

\section{Principles of Event-Based Optical Measurement}
\label{sec:evs}

\subsection{Event generation and temporal encoding}

Event-based vision sensors operate as asynchronous temporal encoders in which individual pixels independently generate events whenever the local variation in logarithmic illumination exceeds a predefined contrast threshold. Unlike conventional frame-based image sensors that periodically sample the entire sensing plane, EVS continuously monitor temporal illumination variations and transmit only brightness-change events. Consequently, the measured output is not a sequence of image frames, but a sparse stream of timestamped events representing threshold-crossing dynamics in time.

Each pixel stores a memorized reference level associated with the last detected brightness state. Events are generated whenever the difference between the instantaneous logarithmic response and the memorized reference exceeds a positive (\(C_{Pos}\)) or negative \(C_{Neg}\)threshold,

\begin{equation}
    e_k =
    \begin{cases}
        (x_k,t_k,+1), & \Delta L(x,t) \geq C_{\mathrm{Pos}}, \\
        (x_k,t_k,-1), & \Delta L(x,t) \leq -C_{\mathrm{Neg}},
    \end{cases}
    \label{eq:event_generation}
\end{equation}
where \(e_k\) denotes the \(k\)-th event, \(x_k\) its pixel location, \(t_k\) its timestamp, and \(\Delta L(x,t)\) the temporal variation in logarithmic illumination relative to the memorized pixel reference state, and the final term denotes event polarity \cite{gallego2020, nozaki2017temperature}. The event stream, therefore, constitutes a nonlinear and asynchronously quantized temporal representation of the optical excitation. In optical measurement applications, this recorded event stream is frequently interpreted as a temporally resolved representation of the underlying optical waveform. However, the temporal fidelity of this reconstruction depends not only on timestamp precision, but also on the physical mechanisms governing event generation, reset dynamics, and readout serialization.

\subsection{Pixel Dynamics and Temporal Response}
The temporal response of the sensor is fundamentally constrained by the dynamics of the photoreceptor and by the threshold-triggered event-generation process. Following an optical transition occurring at \(t=t_0\) with an intensity variation from \(I_0\) to \(I_1\), the photoreceptor does not instantaneously reach its new equilibrium state but rather evolves approximately as:

\begin{equation}
    V_p(t) =
    V_{p,1}
    +
    \left( V_{p,0} - V_{p,1} \right)
    \exp \left[-\frac{t-t_0}{\tau_p}\right]
    \label{eq:photoreceptor_step_response}
\end{equation}
where \(V_{p,0}\) and \(V_{p,1}\) are the initial and final photoreceptor states and \(\tau_p\) is the effective photoreceptor time constant. The value of \(\tau_p\) is dependent on both photocurrent magnitude and pixel bias configuration \cite{hu2021v2e}. In practice, temporal response is faster under more intense illumination conditions due to increased photocurrent-driven charging dynamics, and slower in low-light operation, leading to settling behavior and increased temporal broadening \cite{hu2021v2e}. 

This illumination-dependent relaxation behavior directly influences the event-generation process, since events are emitted whenever the instantaneous logarithmic photoreceptor response exceeds the contrast threshold relative to the memorized reference state. As a result, a single optical transition can produce multiple consecutive events from the same pixel. Following each detected event, the internal reference level is updated to the new threshold-crossing state, whereas the analog photoreceptor signal continues to evolve toward its equilibrium value according to the intrinsic photoreceptor dynamics. Consequently, the residual relaxation of the photoreceptor response can repeatedly drive the signal across successive threshold levels, generating temporally clustered burst-like event sequences associated with a single physical illumination transition.

\begin{figure}[h!]
    \centering
    \includegraphics[width=0.45\textwidth]{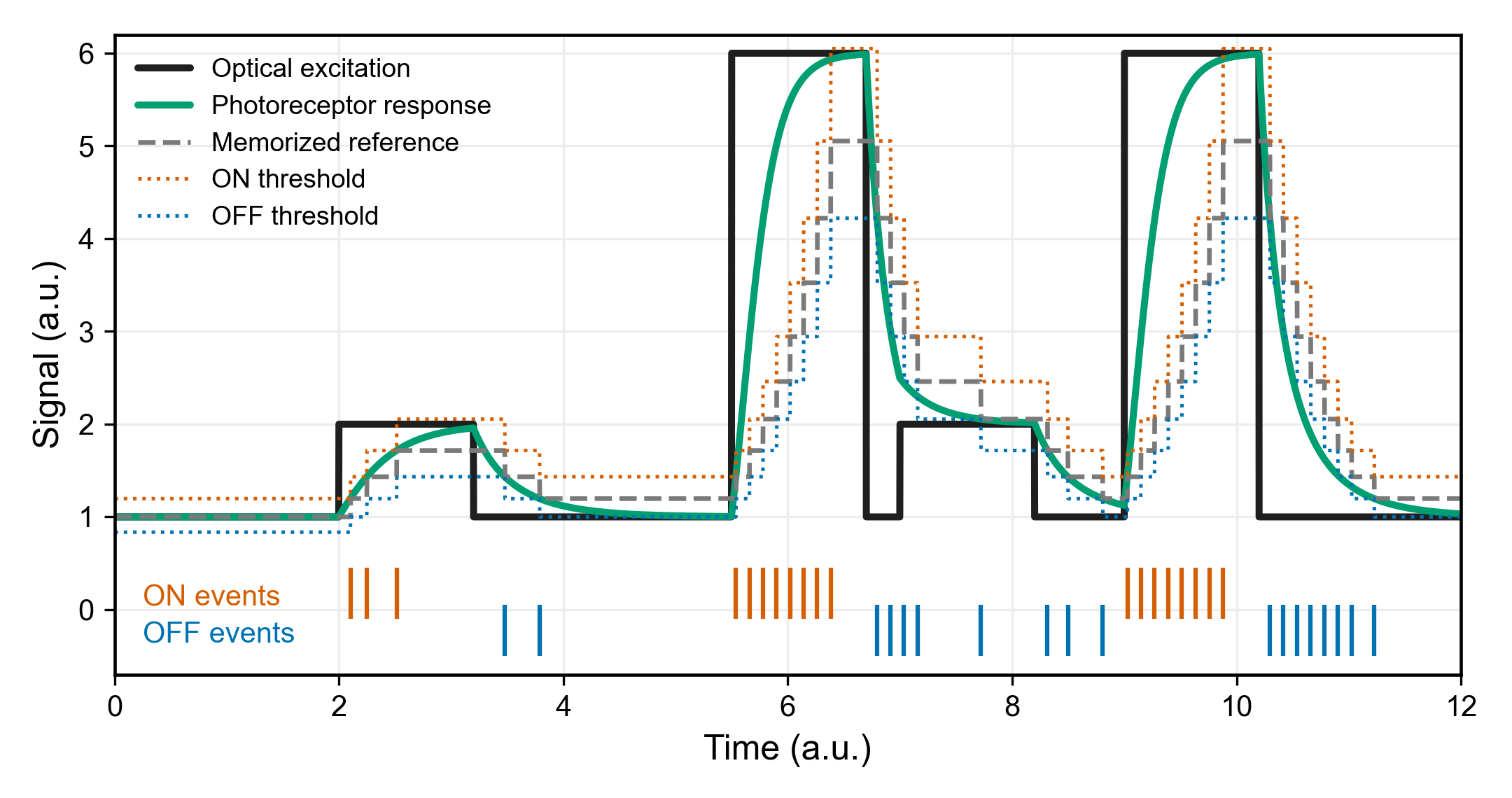}
    \caption{Simulated behavior of the contrast detection signal in relation to the photocurrent in event detection.}
    \label{fig:evs-time}
\end{figure}

The temporal response measured from the event stream is not determined exclusively by the optical excitation itself, but by the combined effect of photoreceptor time constant, contrast threshold, and magnitude of the illumination change. This is particularly important when interpreting temporally varying optical signals. The earliest detectable event defines the transition detectability limit of the sensor, whereas faithful waveform reconstruction depends on the temporal distribution of the complete event population generated during the optical transition. As a result, microsecond-scale event detection does not necessarily imply microsecond-scale fidelity in waveform reconstruction. Under rapidly repeated optical excitation, additional limitations arise from polarity-dependent reset dynamics. If OFF-event generation is delayed or suppressed, the internal pixel reference cannot fully return to the lower illumination state before the next excitation cycle. The effective logarithmic contrast associated with subsequent transitions is therefore progressively reduced, limiting the ability of the sensor to sustain stable event generation at high modulation frequencies, as can be observed in Figure \ref{fig:evs-time}.

\subsection{Readout Serialization and Temporal Dispersion}
After threshold crossing at the pixel level, detected events propagate through the asynchronous sensor readout architecture before timestamp assignment. In the IMX636 sensor, a pixel first generates an analog row request, \(ReqY\), which is processed by an asynchronous row-arbitration stage. Once the selected row is acknowledged, active pixels within that row emit their column requests, \(ReqX_{\mathrm{on}}\) or \(ReqX_{\mathrm{off}}\), according to event polarity \cite{prophesee_csd4mhdcd_2020,prophesee_readout_precision}. The row address and column-event vectors are subsequently transferred to the synchronous readout stage, where timestamps are assigned. Thus, the timestamp associated with an event corresponds to the time at which the event reaches this readout stage, not exactly to the instant at which the pixel comparator first detected the contrast crossing. This distinction is essential for the interpretation of time-resolved optical measurements.

For isolated events, Prophesee reports a typical delay of approximately \(60\)--\(70\,\mathrm{ns}\) between pixel-level event generation and timestamp assignment. Consecutive row selections occur approximately with a delay of \(120\,\mathrm{ns}\), comprised of about \(60\,\mathrm{ns}\) to create the timestamp and \(60\,\mathrm{ns}\) to prepare the next row selection \cite{prophesee_csd4mhdcd_2020, prophesee_readout_precision}. Events generated simultaneously across different rows are therefore inherently serialized by the arbitration process and are not necessarily assigned identical timestamps.

After row selection, active pixels are transferred via a vectorized column-readout architecture, in which each row is partitioned into 40 vectors, each with 32 columns. Only vectors containing active pixels are transmitted. Consequently, the row-transfer time depends on the spatial distribution of activity, ranging from approximately \(60\, \mathrm{ns}\) when a single vector is occupied to approximately \(460\, \mathrm{ns}\) when all 40 vectors of a 1280-pixel row contain activity \cite{prophesee_csd4mhdcd_2020, prophesee_readout_precision}. As a result, the effective readout-induced temporal spreading depends on the size and spatial distribution of the active region during acquisition. 

Under sparse excitation involving only a few active rows and column vectors, readout-induced temporal dispersion remains comparatively limited and typically occurs on sub-microsecond timescales. In contrast, under spatially extended or near-global optical excitation regime can simultaneously activate a large number of rows and column vectors, and measured event stream can become temporally broadened over substantially longer intervals despite nearly simultaneous physical threshold crossings at the pixel level. The measured temporal response must therefore be interpreted not only as a property of the pixel-level contrast detector, but also as a consequence of the spatial extent of the excitation and of the asynchronous readout architecture itself.

This point is particularly relevant for scientific applications, where event-based sensors are increasingly used as high-speed optical detectors, but where the internal detection and readout mechanisms are not always fully accounted for. In many applications, the relevant signal is not a sparse moving edge but a spatially extended change in illumination, for example, a global pulse, a near-global modulation, or a stimulus for which the relative timing between different pixels must be compared. In such cases, finite pixel response, polarity dependence, threshold-crossing dynamics, refractory behavior, and readout-induced temporal spreading collectively determine the temporal fidelity of the measured event stream.

\begin{figure*}[h!]
    \centering
    \includegraphics[width=0.9\linewidth]{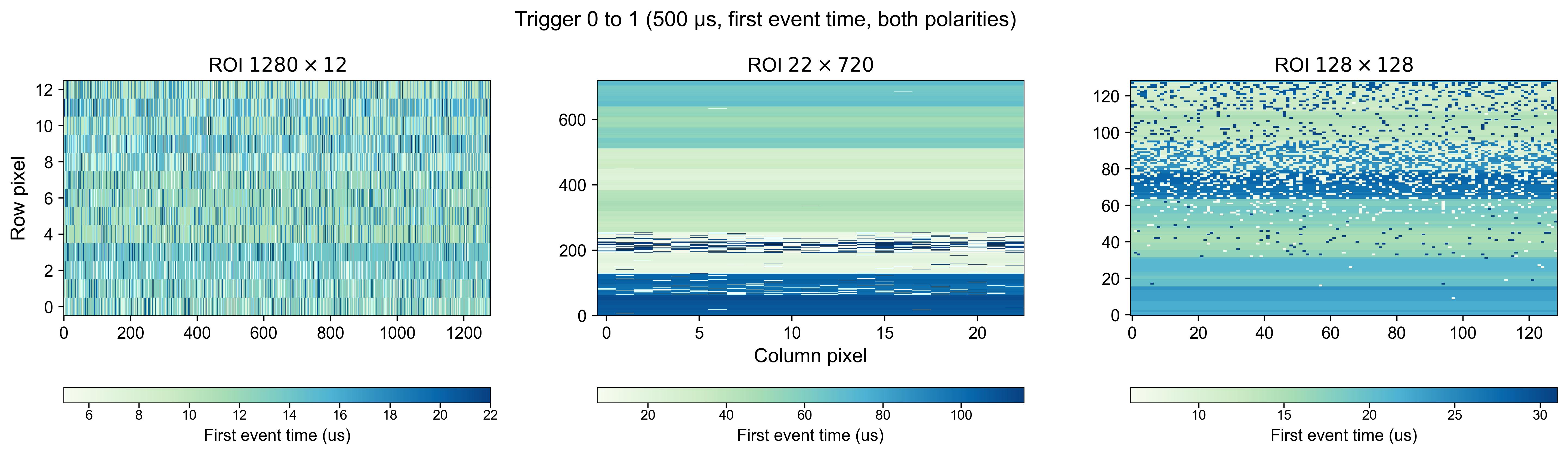}
    \caption{
        Measured first-event timing maps for three ROI geometries following a pulsed optical transition (LED turned on for \SI{500}{\micro\second}). Each pixel is colored according to the timestamp of its first event within the trigger window, measured relative to the first trigger. The three panels correspond to ROIs of \(1280 \times 12\), \(22 \times 720\), and \(128 \times 128\) pixels. Although the illumination transition is spatially extended, the events are not reported simultaneously across the active region. The vertically extended ROI shows a stronger row-dependent timing structure because activity is distributed across many sensor rows, increasing the contribution of row arbitration and readout serialization. In contrast, the horizontally extended ROI is confined to fewer rows and exhibits a narrower first-event timing range. These measurements illustrate how ROI geometry contributes to readout-induced temporal dispersion in the event stream.
        }
    \label{fig:Serialization}
\end{figure*}

Figure \ref{fig:Serialization} illustrates this effect using measured first-event timing maps obtained after a pulsed optical transition for three ROI geometries. Each pixel is colored according to the timestamp of its first event within the trigger window. Although the optical excitation is applied globally, the measured response is not reported simultaneously across the active region. Instead, the timing structure depends on the spatial extent of the ROI and on the row-based arbitration and readout serialization of the sensor. The vertically extended ROI exhibits a stronger row-dependent timing gradient. In contrast, the horizontally extended ROI remains confined to a smaller range of rows and therefore shows reduced readout-induced temporal spreading.

A dedicated characterization is therefore required to determine how the camera responds under controlled optical excitation. In this work, we perform this characterization using sinusoidal and pulsed illumination. Sinusoidal illumination is used to evaluate the frequency response of the sensor and to quantify how the event stream tracks periodic changes in light intensity. Pulsed illumination is used to study the time-domain response to sharp transitions, the recovery dynamics under rapidly repeated excitation, and the fidelity with which short optical pulses can be reconstructed from the event stream. Together, these measurements provide a practical characterization of the IMX636 response for applications in which event cameras are used not only for vision, but also as time-resolved optical sensing instruments.

\section{Experimental Setup}
\begin{figure*}[h!]
    \centering
    \includegraphics[width=0.7\linewidth]{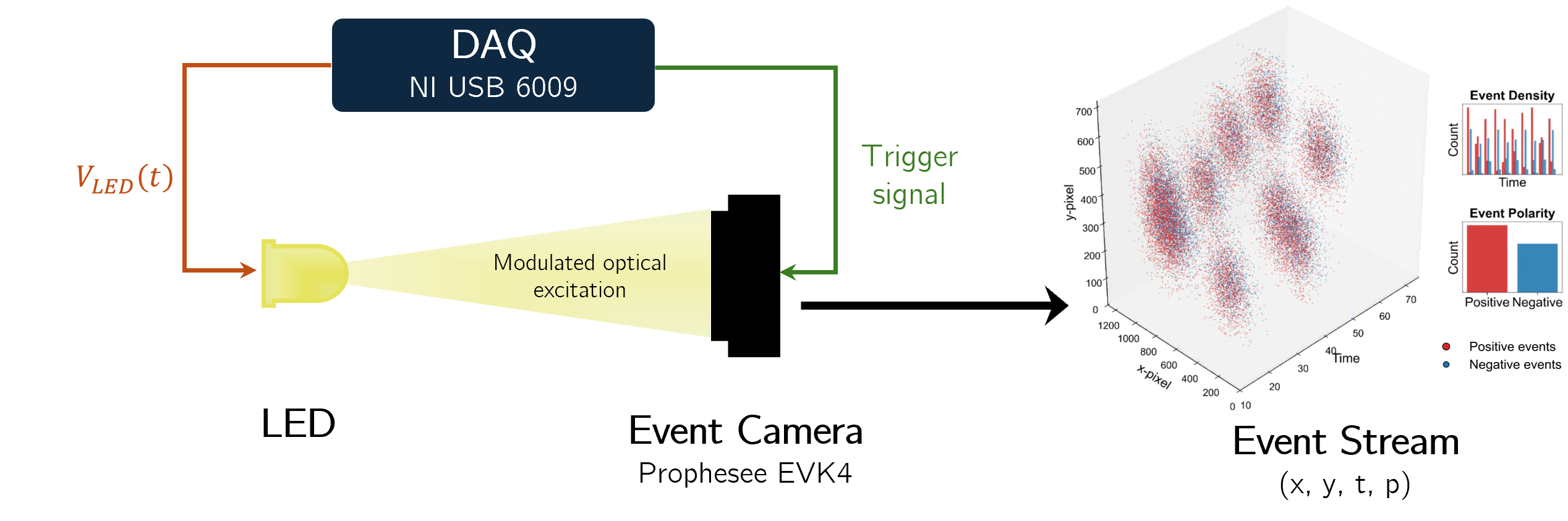}
    \caption{The setup used for characterization comprised an LED pointed at the EVS camera sensor. The light is modulated using a National Instruments DAQ, with trigger pulses being sent to the camera to signal optical pulse transitions.}
    \label{fig:setup}
\end{figure*}

To characterize the temporal response of the event-based sensor under controlled optical excitation, the experimental setup shown in Figure \ref{fig:setup} was developed. Optical stimuli were generated using a National Instruments USB-6009 data acquisition device (DAQ), which produced both the modulation waveform applied to the light-emitting diode (LED) and a synchronized trigger signal for the event camera. The LED converted the electrical excitation into controlled variations in optical intensity, which were recorded by an IMX636-based event camera (Prophesee EVK4). The DAQ-generated waveform was used to drive the LED directly, enabling precise control of the illumination temporal profile. Simultaneously, a trigger signal from the DAQ was supplied to the event camera to synchronize the optical excitation with the recorded event stream. Event data were acquired through the Metavision framework and subsequently processed offline.

For frequency-domain characterization, the LED intensity was modulated using sinusoidal waveforms with adjustable offset voltage, peak-to-peak amplitude, and modulation frequency. The modulation frequency was varied over the range investigated in this work to evaluate the temporal response of the sensor under periodic optical excitation. These measurements were used to characterize event rate, signal-to-noise ratio, lock-in amplitude, polarity balance, and waveform reconstruction fidelity.

For time-domain characterization, pulsed optical excitation was generated by applying voltage pulses of controlled width and repetition rate to the LED. These measurements were used to investigate first-event latency, response duration, recovery dynamics, and pulse reconstruction fidelity under rapidly varying optical excitation.

\section{Frequency Characterization}
\label{subsec:frequency_characterization}

The frequency-domain response of an event camera is fundamentally determined by the interaction between optical excitation dynamics, pixel-level event-generation mechanisms, and the asynchronous readout architecture. Although event cameras provide microsecond timestamp resolution, the ability of the recorded event stream to faithfully represent a rapidly varying optical signal depends on several factors, including photoreceptor temporal response, threshold-crossing dynamics, refractory behavior, polarity balance, and readout-induced temporal dispersion. The objective of this characterization is therefore not only to determine the maximum detectable modulation frequency, but also to identify the mechanisms responsible for the degradation of temporal fidelity at high frequencies. Periodic optical excitation was generated by driving the LED with

\begin{equation}
    V_{\mathrm{LED}}(t) = V_{\mathrm{offset}} - \frac{V_{\mathrm{pp}}}{2} \cos(2\pi f t),
\end{equation}
where \(V_{\mathrm{offset}}\) controls the average LED drive level, and therefore the average baseline illumination reaching the sensor, while \(V_{\mathrm{pp}}\) defines the peak-to-peak modulation amplitude. For this characterization, the modulation frequency \(f\) was swept logarithmically from \(1\,\mathrm{kHz}\) to \(350\,\mathrm{kHz}\). 

To minimize event-rate limitations imposed by the camera configuration, the refractory bias was set to the maximum value supported by the IMX636 sensor \((bias_{\mathrm{refr}}=235)\), corresponding to a nominal minimum pixel dead time of approximately \SI{19}{\micro\second} according to the Metavision SDK. Although this value does not directly define the optical bandwidth of the sensor, it establishes an important characteristic timescale for interpreting the response under high-frequency excitation.

The response was quantified using four complementary metrics. Local signal-to-noise ratio (SNR) and lock-in amplitude were used to evaluate the strength and spectral coherence of the response at the imposed modulation frequency. In parallel, the polarity balance and event rate were used to characterize the event-generation process itself, allowing changes in frequency response to be associated with variations in event activity, polarity asymmetry, or loss of synchronization with the optical stimulus.

The local SNR quantifies how clearly the imposed modulation frequency stands out against the surrounding spectral background. To prevent broad low-frequency drift from skewing the metric, the signal-to-noise ratio is calculated locally using the peak power near the target frequency (\(P_{\mathrm{peak}}\)) and the median power of the surrounding noise band (\(\tilde{P}_{\mathrm{noise}}\)):
    \begin{equation}
        \mathrm{SNR}_{\mathrm{dB}} = 10 \log_{10} \left( \frac{P_{\mathrm{peak}} - \tilde{P}_{\mathrm{noise}}}{\tilde{P}_{\mathrm{noise}}} \right).
    \end{equation}

The lock-in amplitude, which measures the strength of the periodic component of the event response exactly at the commanded modulation frequency \(f\), is equivalent to a single-frequency digital lock-in detector. For a windowed, mean-centered discrete event signal \(s_n\), sampled at times \(t_n\) with window weights \(w_n\), the amplitude \(A_f\) is given by:
    \begin{equation}
        A_f = \frac{2}{\sum_n w_n} \left| \sum_{n} s_n w_n e^{-i 2\pi f t_n} \right|.
    \end{equation}

To characterize the event-generation dynamics, the polarity balance was defined as:
    \begin{equation}
        B = \frac{N_{\mathrm{Pos}} - N_{\mathrm{Neg}}}{N_{\mathrm{Pos}} + N_{\mathrm{Neg}}}.
    \end{equation}
     And it quantifies the relative distribution of positive (\(N_{\mathrm{Pos}}\)) and negative (\(N_{\mathrm{Neg}}\)) events generated during the acquisition: Values near \(+1\) indicate a dominance of positive events, values near \(-1\) indicate a dominance of negative events, and values near zero indicate balanced detection.

Finally, the overall event activity was quantified through the event rate,
    \begin{equation}
        R = \frac{N_{\mathrm{Pos}} + N_{\mathrm{Neg}}}{T},
    \end{equation}
where \(T\) is the acquisition duration.

Two complementary experiments were performed. In the first, the modulation amplitude was fixed while the offset voltage was varied to investigate the influence of the background average illumination level. In the second experiment, the offset voltage was fixed near the optimal operating point identified in the first experiment, while the modulation amplitude was varied. Measurements were performed for multiple ROI geometries, such as 1280x12, 22x720, and 128x128, to evaluate the influence of spatial activity on the temporal response. In the main manuscript, we present results for the 128x128 case; results for the other geometries are available in the supplementary material.

\subsection{Background illumination characterization}
\begin{figure*}[h!]
    \centering
    \includegraphics[width=0.9\linewidth]{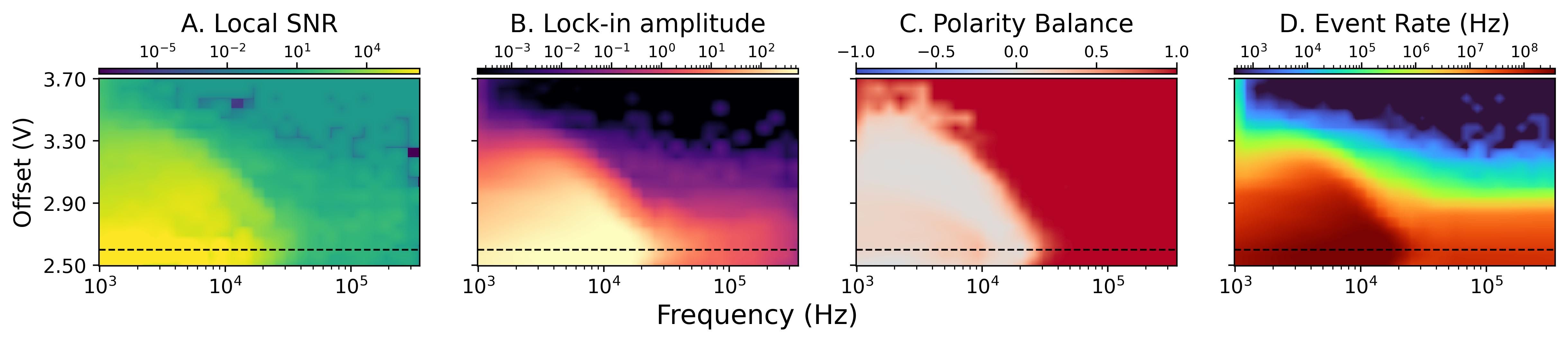}
    \caption{Frequency-response characterization of the IMX636 sensor as a function of LED offset voltage for a ROI of 128 × 128 pixels, using a fixed modulation amplitude of \(V_{pp}=0.2\,\mathrm{V}\). The refractory bias was fixed to 235 to minimize the pixel dead time. (A) Local SNR, (B) lock-in amplitude, (C) polarity balance, and (D) event rate as functions of modulation frequency and LED offset voltage. Increasing the offset initially extends the detectable frequency range by increasing event activity, whereas larger offsets progressively suppress negative event generation, leading to reduced waveform fidelity despite the higher event rate. The dashed black line marks the 2.6\,V operating point.}
    \label{fig:I0}
\end{figure*}

The first experiment investigated the effect of the average illumination level on the frequency response of the sensor. The modulation amplitude was fixed at $V_{pp}$ = 0.2\,V, while the LED offset voltage (\(V_\mathrm{offset}\)) was swept from 2.5\,V to 3.7\,V in 20 equally spaced steps. The choice of 2.5\,V was done empirically, approximating the lowest operating voltage of the LED. Figure~\ref{fig:I0} presents the resulting frequency-response maps for a selected ROI of 128 \(\times\) 128 pixels, while the corresponding results for the remaining ROI geometries are provided in section S1 of the Supplementary Material.

We start the analysis by noticing that for the lowest offset voltage value considered, the average optical power reaching the sensor may be insufficient to generate a large population of events. Consequently, the event rate may remain low, and both the local SNR and lock-in amplitude rapidly deteriorate as the modulation frequency increases. Increasing the offset voltage initially improves the response by increasing the baseline illumination and, therefore, the number of generated events. This results in a clear extension of the detectable frequency range, accompanied by higher lock-in amplitudes and improved local SNR. However, this improvement is not monotonic. Beyond an offset voltage of approximately \(2.6\,\mathrm{V}\), the frequency response progressively degrades despite the continued increase in average illumination. This behavior is accompanied by a marked shift in the polarity balance toward positive values, indicating that the generation of negative events becomes increasingly suppressed. Since accurate reconstruction of a periodic optical signal requires balanced positive and negative responses, the loss of negative events reduces waveform fidelity and leads to the observed reduction in local SNR.

The event-rate maps help clarify what is happening during this transition. Although the overall event activity continues to increase with offset voltage, the corresponding improvement in frequency response saturates. The simultaneous increase in event rate and deterioration in polarity balance indicates that the limiting factor is no longer the availability of events, but rather the ability of the sensor to generate temporally balanced responses during successive illumination cycles. 

Based on these observations, an offset voltage of \(2.6\,\mathrm{V}\) was selected for the remainder of the frequency-domain characterization, providing the best compromise between event activity, polarity balance, and temporal response.

\subsection{Modulation amplitude characterization}

\begin{figure*}[h!]
    \centering
    \includegraphics[width=0.9\linewidth]{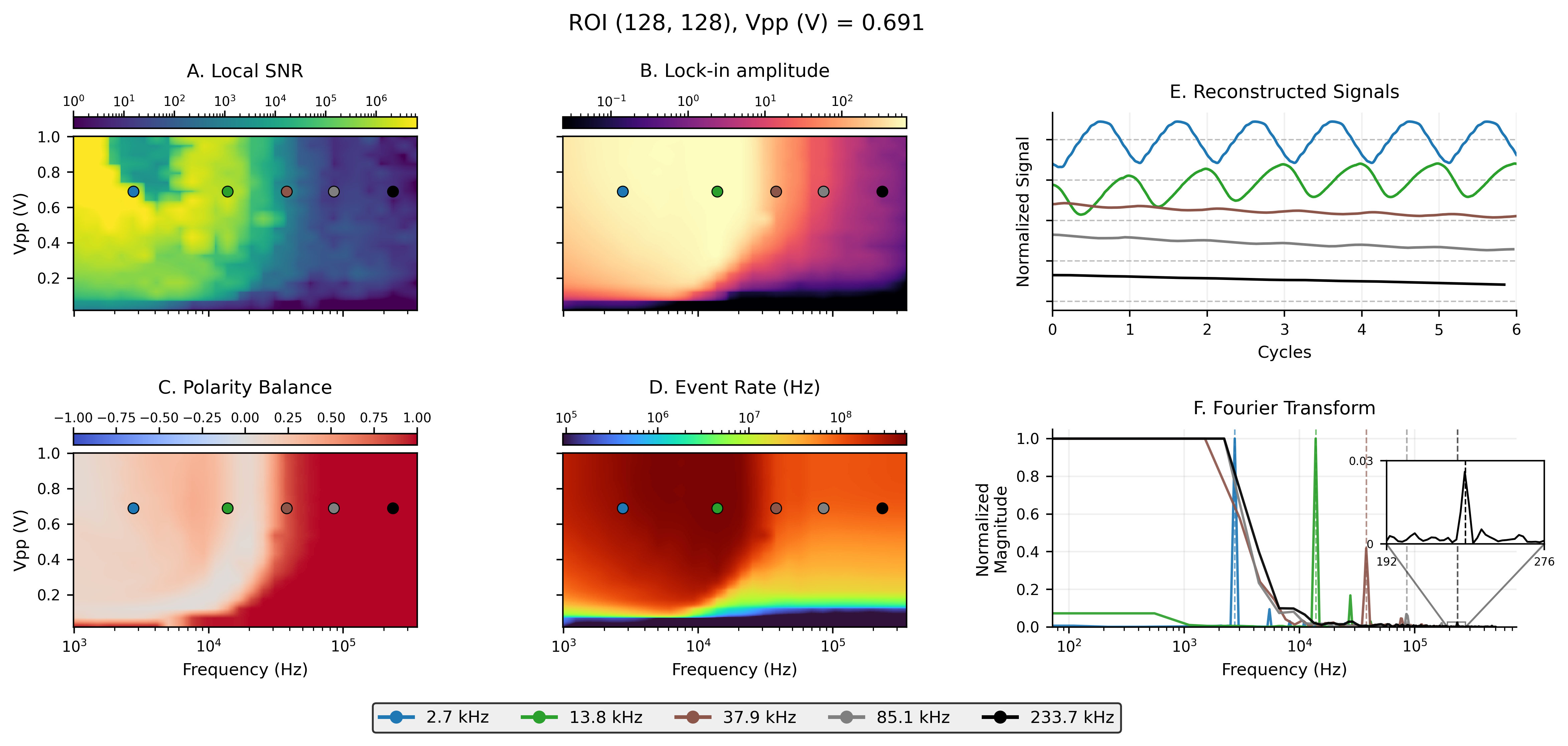}
    \caption{
    Frequency-response characterization of the IMX636 sensor as a function of modulation amplitude \(V_{pp}\), with the illumination offset fixed near the optimal operating point identified in Figure~\ref{fig:I0} for a ROI of 128 \(\times\) 128 pixels. (A) local SNR at the target frequency; (B) lock-in amplitude; (C) polarity balance; (D) event rate. The colored markers indicate the selected operating points used for the time-domain reconstruction and Fourier analysis in (E) and (F), respectively; (E) Reconstructed signals obtained from the cumulative signed event stream, shown over six modulation cycles and vertically offset for clarity; (F) Corresponding normalized Fourier spectra computed from the same signal segments. Dashed vertical lines indicate the commanded modulation frequencies, and the insets zoom into the highest-frequency selected point, where the spectral peak is weak but still visible.
    }
    \label{fig:deltaI}
\end{figure*}

Still on the evaluation of the frequency response of the IMX636 sensor, a second experiment was subsequently performed to evaluate the influence of modulation amplitude (\(V_{\mathrm{pp}}\)). The LED offset (\(V_\mathrm{offset}\)), was fixed at \(2.6\,\mathrm{V}\), while the peak-to-peak voltage was varied from \(0.02\,\mathrm{V}\) to \(1\,\mathrm{V}\) using 20 equally spaced values. Figure~\ref{fig:deltaI} presents the resulting frequency-response maps for the 128 \(\times\) 128 ROI, while the corresponding results for the remaining ROI geometries are provided in section S2 of the Supplementary Material.

At low modulation amplitudes, the optical contrast is only marginally above the event threshold, resulting in sparse event generation and weak synchronization with the optical excitation. Consequently, both the local SNR and the lock-in amplitude decrease rapidly as the modulation frequency increases, thereby limiting the detectable frequency range. As the modulation amplitude increases, larger logarithmic intensity variations produce more reliable threshold crossings, increasing the number of generated events and extending the high-SNR region toward higher modulation frequencies. Systematic increases in both the lock-in amplitude and the event rate accompany this improvement. However, the improvement gradually saturates as the modulation amplitude continues to increase. Although larger amplitudes continue to generate more events, the maximum detectable frequency exhibits only modest gains. Simultaneously, the polarity balance progressively shifts toward positive values at high frequencies, indicating an increasing suppression of negative events. Consequently, additional optical contrast no longer translates into proportional improvements in waveform fidelity.

These observations reveal two distinct operating regimes. At low modulation amplitudes, the response is contrast-limited and governed primarily by the ability of the optical signal to exceed the event threshold. Once sufficient contrast has been achieved, the response becomes increasingly dynamics-limited, with the recoverable bandwidth determined by the intrinsic temporal behavior of the sensor, including refractory dynamics, photoreceptor response, and asynchronous event readout.

To illustrate this transition, representative operating points spanning the measured frequency range were selected and are shown in Figure~\ref{fig:deltaI}. For frequencies up to 15kHz, the reconstructed signals (Figure~\ref{fig:deltaI}E) exhibit stable periodic oscillations at low modulation frequencies, where positive and negative events are generated in a balanced manner and accurately reproduce the imposed optical waveform. As the modulation frequency increases, the reconstructed waveforms progressively lose temporal contrast and become increasingly asymmetric, reflecting the gradual suppression of negative events already observed in the polarity-balance maps. This is also confirmed by the corresponding Fourier spectra, presented in (Figure~\ref{fig:deltaI}F). At low frequencies, the commanded modulation frequency appears as a narrow and well-defined spectral peak with minimal surrounding background. As the modulation frequency increases, the target-frequency component remains clearly identifiable, but becomes progressively embedded within the surrounding spectral noise floor. Consequently, the response at the target frequency persists beyond the apparent local-SNR cutoff, even though the reconstructed waveform has already undergone significant degradation.

The remaining ROI geometries exhibit the same qualitative behavior (see Supplementary material), with differences in the maximum recoverable frequency as discussed in the following subsection.

\subsection{Influence of ROI Geometry and Pixel-Level Dynamics}
The results presented in the previous sections demonstrate that the recoverable bandwidth depends on the illumination conditions and modulation parameters. While different ROI geometries were considered in the analysis, the qualitative trends with illumination level and modulation amplitude remain consistent across configurations. 

\begin{figure*}[h!]
    \centering
    \includegraphics[width=0.75\linewidth]{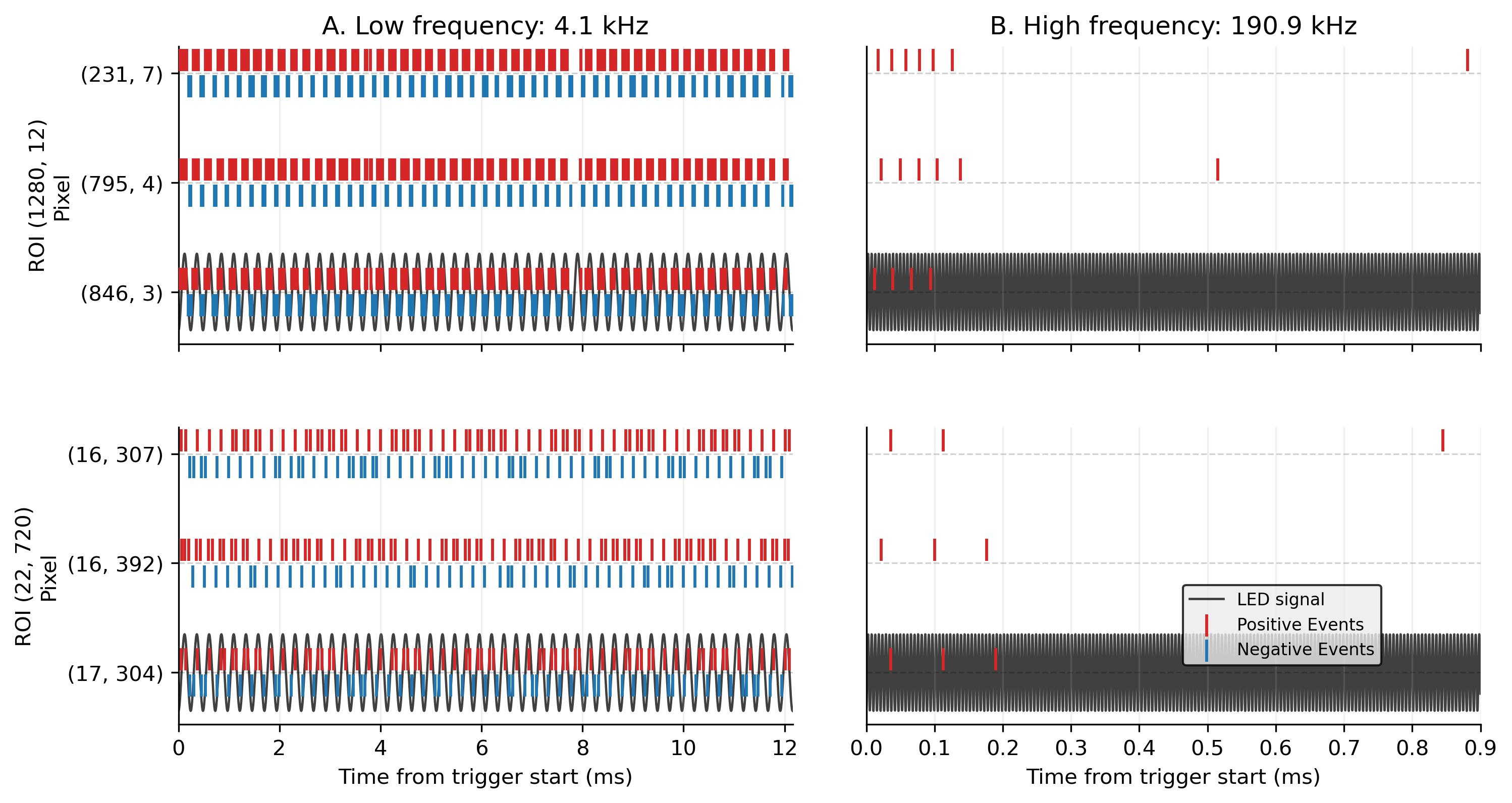}
    \caption{
    Pixel-level event timestamp comparison for representative low- and high-frequency LED modulation conditions. Each panel shows 3 selected pixels (vertical axis) within the corresponding ROI, with positive and negative events plotted as timestamp markers and the imposed LED modulation shown for reference in the pixel shown at the bottom. The comparison highlights how individual pixels can follow the low-frequency modulation more consistently when the ROI is chosen with fewer rows, whereas at high frequencies, the event response becomes sparse and temporally irregular due to refractory and readout-related limitations.
    }
    \label{fig:pixel_response}
\end{figure*}

To verify if the aggregate frequency response corresponds to true single-pixel tracking or instead emerges from the summed response of many sparsely firing pixels, representative pixels were selected from the horizontal \((1280 \times 12)\) and vertical \((22 \times 720)\) ROI geometries and their event timestamps were plotted directly, this can be found in Figure \ref{fig:pixel_response}. This analysis provides a bridge between the ROI-level spectral metrics and the underlying event-generation process. In particular, it allows us to distinguish between two regimes: one in which individual pixels repeatedly fire throughout the stimulus window, and another in which the reconstructed frequency component is produced mainly by the temporal distribution of events across the full population of pixels.

For the horizontal ROI - i.e., lower number of rows - the low-frequency excitation produces regular sequences of positive and negative events throughout the acquisition window, with all monitored pixels remaining active from the beginning to the end of the recording. At a higher frequency of \(190.9,\mathrm{kHz}\), however, activity becomes concentrated during the initial cycles of the stimulus. Only \(0.78\%\) of the monitored pixels remain active during the final \(\SI{200}{\micro\second}\) of the acquisition, the most active pixels generate eleven events, and no negative events are observed. Therefore, at this high frequency, the ROI-level periodic response does not arise from individual pixels tracking every modulation cycle, but from the aggregate contribution of many pixels firing intermittently throughout the acquisition. The absence of negative events can be explained by the polarity-dependent reset dynamics of the sensor. Each positive event updates the internal reference level to a higher illumination value. When negative-event generation becomes delayed or suppressed, the reference level is not efficiently restored during the falling portion of the optical cycle. Consequently, subsequent positive events can only be generated over a progressively smaller fraction of the modulation period, producing the observed transition from regular periodic firing to sparse and apparently irregular activity.

The vertical ROI - i.e., a larger number of rows - exhibits the same polarity-dependent behavior but with an additional limitation imposed by the readout geometry. Because activity is distributed across 720 sensor rows, row arbitration introduces additional temporal spreading. Individual pixels exhibit inter-event intervals of approximately 77-\SI{80}{\micro\second}, comparable to the expected accumulated row-selection time for this geometry. At \(190.9,\mathrm{kHz}\), only \(0.32\%\) of monitored pixels remain active during the final \SI{200}{\micro\second} of the acquisition, and the most active pixels generate only four events. The combined effects of polarity imbalance and readout serialization, therefore, explain the lower recoverable bandwidth observed for this ROI.

These observations demonstrate that the effective frequency response of an event camera cannot be interpreted as the temporal bandwidth of an individual pixel. Instead, it emerges from the interaction between pixel-level dynamics, polarity-dependent event generation, refractory behavior, and the asynchronous readout architecture. Consequently, ROI selection should be regarded not only as a spatial sampling parameter but also as a temporal design configuration when event cameras are used for quantitative optical measurements. Note that the frequency limits discussed here are specific to the minimum-refractory-time configuration used in this work. Other refractory-bias settings are expected to shift the effective cutoff, with longer refractory times reducing the usable high-frequency range. However, the same qualitative behavior should remain: increasing modulation amplitude improves detection only up to the point where refractory dynamics, polarity imbalance, photoreceptor response, and readout serialization become the dominant limitations.

\section{Pulse Characterization}
\label{sec:pulse_char}

The frequency-domain characterization presented in the previous section quantifies the ability of the IMX636 sensor to track continuous optical modulation. However, it does not fully capture the transient response of the sensor to abrupt changes in illumination. To investigate these dynamics, a series of pulse-based experiments was performed, focusing on response latency, recovery dynamics, and temporal reconstruction fidelity.

Three pulse-based measurements were performed. First, isolated optical pulses were used to characterize the response to individual illumination transitions. Second, short-pulse trains with variable pulse separation were used to evaluate recovery dynamics under repeated excitation. Third, isolated pulses with variable width were used to assess the fidelity with which optical pulse widths can be reconstructed from the event stream.

\subsection{Isolated Step-Response Characterization}
\label{subsec:isolated_step_response}

The first pulse experiment was designed to characterize the response of the IMX636 sensor to a single optical transition. The LED was driven with pulses of \(1\,\mathrm{ms}\) duration, separated by \(20\,\mathrm{ms}\). This separation is significantly longer than the expected sensor response and readout times, ensuring that the event bursts generated by successive pulses do not overlap and can therefore be analyzed independently.
edges
For each region of interest, event timestamps were referenced to the corresponding trigger edge. Rising transitions were analyzed using positive events, while falling edge were analyzed with events with negative polarity. Let \(\mathcal{P}_p\) denote the set of pixels generating at least one event of polarity \(p \in \{+1,-1\}\) within the analysis window. For each pixel \(i \in \mathcal{P}_p\), the first and last event timestamps are denoted by \(t_{i,1}^{(p)}\) and \(t_{i,n_i}^{(p)}\), respectively.

Three temporal metrics were extracted to characterize different aspects of the sensor response. The earliest detectable response was quantified by the first-event latency, defined as \(T_{\mathrm{first}}^{(p)} = \min_{i \in \mathcal{P}_p} t_{i,1}^{(p)}\), which corresponds to the earliest event generated by any responding pixel following the optical transition. The temporal spread of the first response across the active region was quantified by \(T_{\mathrm{first,all}}^{(p)} = \max_{i \in \mathcal{P}_p} t_{i,1}^{(p)}\), representing the time required for all responding pixels to generate their first event. Finally, the complete response duration was quantified by \(T_{\mathrm{all}}^{(p)} = \max_{i \in \mathcal{P}_p} t_{i,n_i}^{(p)}\), corresponding to the latest event associated with the transition and therefore providing a measure of the total temporal extent of the event burst. Together, these metrics distinguish between the earliest detectable response, the spatial variability of the response onset across the sensor, and the overall duration of the event-generation process.

\begin{figure}[h!]
    \centering
    \includegraphics[width=0.85\linewidth]{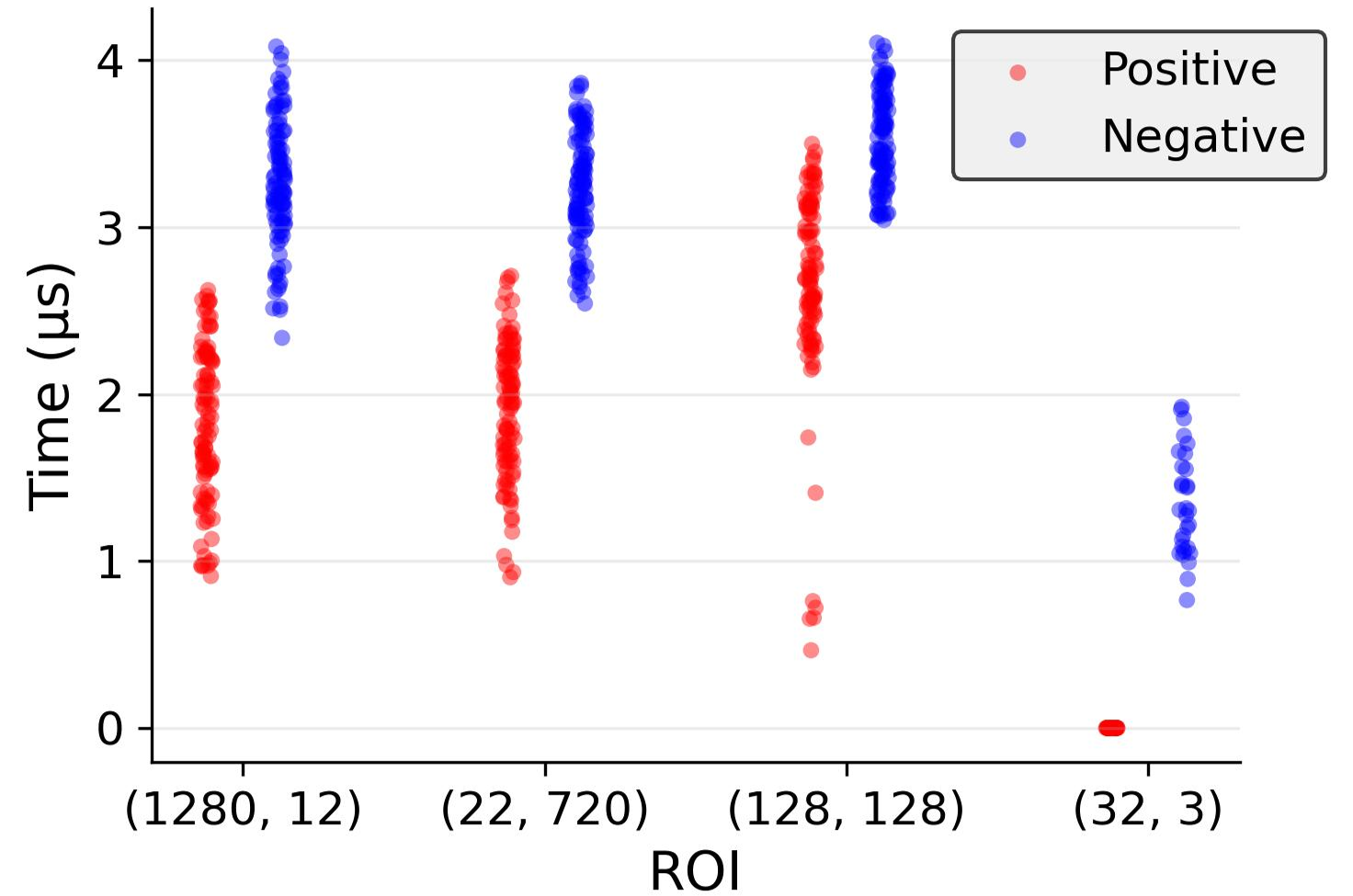}
    \caption{Latency (\(T_{\mathrm{first}}^{(p)}\)) of event detection for each region of interest. The latency is the time elapsed from the trigger that marks the start of the light-increase event to the first timestamp at which an event is recorded.}
    \label{fig:lantency}
\end{figure}

\begin{figure*}[h!]
    \centering
    \includegraphics[width=0.8\textwidth]{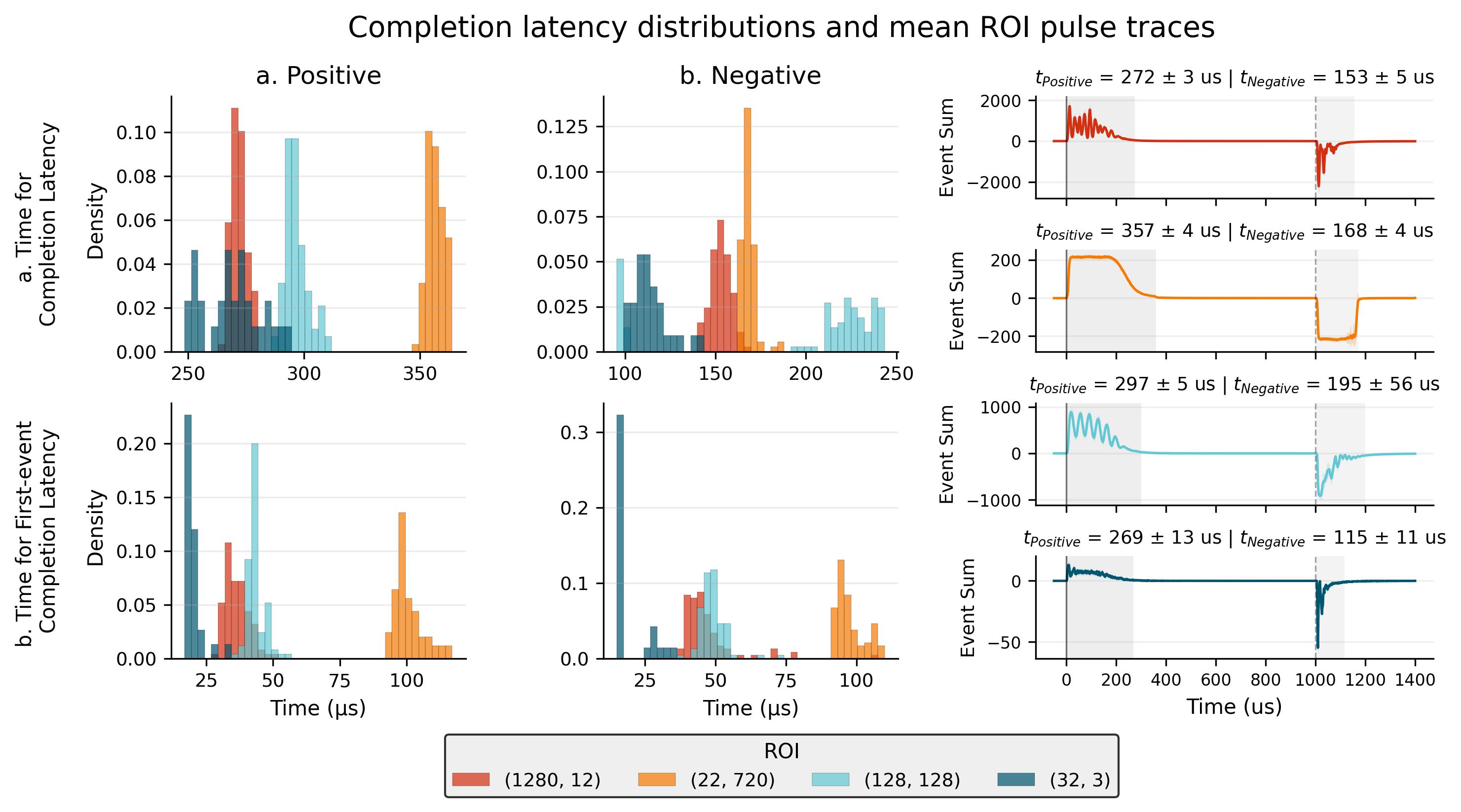}
    \caption{Pulse-response characterization for isolated \(1\,\mathrm{ms}\) optical pulses separated by \(20\,\mathrm{ms}\). The top histograms show the full burst completion time (\(T_{\mathrm{all}}^{(p)}\)), defined as the time required for all events associated with a single rising or falling optical transition to be output. At the bottom, we can see the time required for all responding pixels in the ROI to generate their first event (\(T_{\mathrm{first,all}}^{(p)}\)). The plots on the right show the total number of events generated at the microsecond level, with noticeable differences in readout behavior between ROI geometries.}
    \label{fig:pulse_completion}
\end{figure*}

Figure~\ref{fig:lantency} shows the latency between the optical trigger and the first detected event for the different regions of interest. For all tested geometries, the first event is detected within a few microseconds after the optical transition. Here, we observe that the first-event latency remains weakly dependent on ROI geometry, indicating that the earliest response is dominated by pixel-level contrast detection and the first-row access to the readout pipeline rather than by the total number of active pixels. In other words, even when many pixels are activated, the camera can still report the onset of the optical transition on a microsecond timescale. However, the first-event latency does not reflect the time required for the sensor to fully report the optical step response. This distinction becomes clear in Fig.~\ref{fig:pulse_completion}, where the completion times are shown for the distinct ROIs. As expected from the row-based arbitration in the IMX636 readout architecture, when the optical transition activates many rows nearly simultaneously, the readout serializes the row requests, delaying the first event from all pixels, which causes the time for all active pixels to fire at least once to increase strongly with the vertical extent of the ROI. This means that the time required for the pulse response across the full active region is directly constrained by the spatial geometry of the excitation.

The results demonstrate that the temporal response of an event camera cannot be adequately described by a single latency value. Instead, an isolated optical transition produces a sequence of temporally distinct processes, ranging from the earliest detectable event to the completion of the signal due to the full event burst. This distinction is particularly important for transient optical measurements, where the relevant temporal response depends on the application. While event detection is governed by the first-event latency, faithful reconstruction of rapidly varying optical signals depends on the duration and temporal distribution of the complete event burst.

\subsection{Recovery Dynamics}
\label{subsec:short_pulse_recovery}
To investigate the recovery dynamics of the event-generation process, the LED was driven with trains of short optical pulses while the temporal separation between consecutive pulses was progressively reduced. For this purpose, the LED was driven using pulses of \SI{2.5}{\micro\second}, while the temporal separation between consecutive pulses, \(\Delta t\), was varied. By reducing \(\Delta t\), the experiment explores the effect of incomplete recovery on the event-generation process. For each value of \(\Delta t\), the positive and negative event counts, polarity balance, and event rate were evaluated throughout the pulse train to quantify the evolution of the sensor response as recovery became progressively limited.

\begin{figure*}
    \centering
    \includegraphics[width=0.75\linewidth]{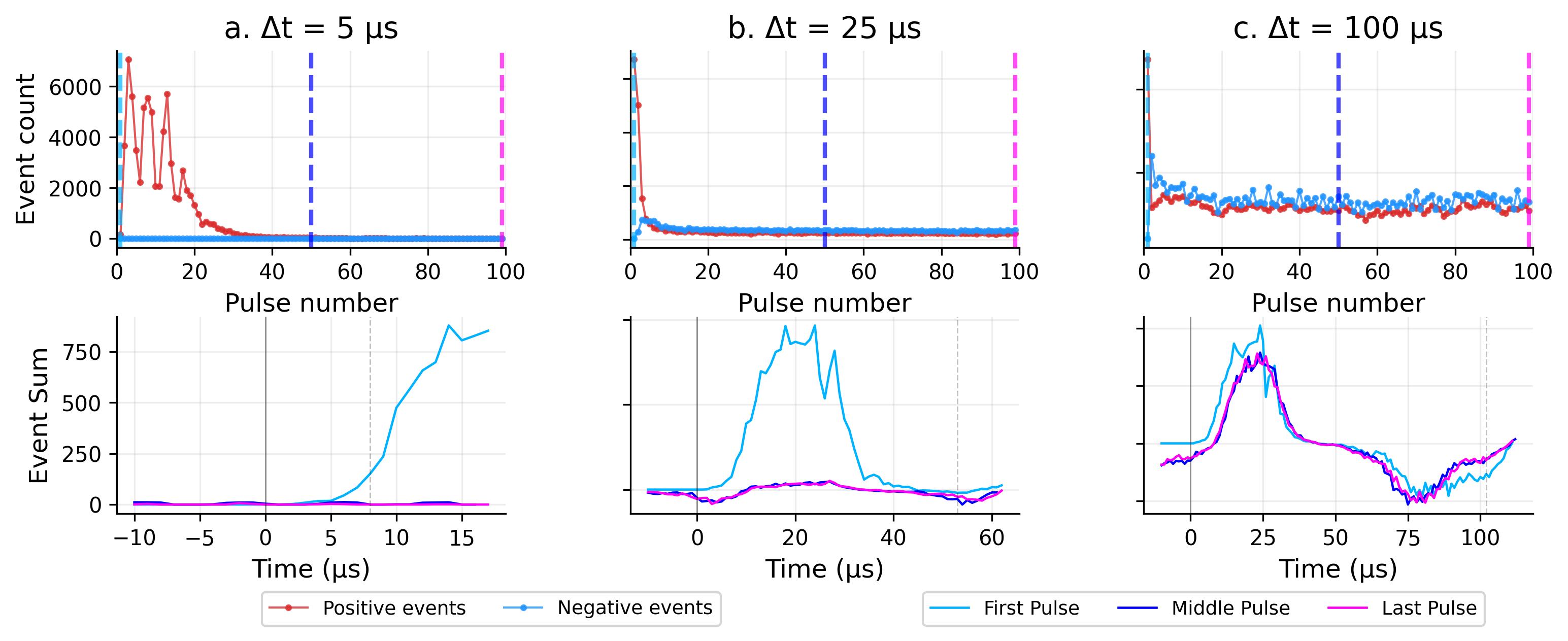}
    \caption{Recovery dynamics under rapidly repeated short-pulse excitation. The LED was driven with \SI{2.5}{\micro\second} pulses while the separation \(\Delta t\) between consecutive pulses was varied. For small \(\Delta t\), events from the rising edge are still being output after the optical pulse has ended, while the delayed or suppressed negative response prevents efficient reset of the internal pixel reference. This results in a progressive loss of event activity throughout the pulse train. For larger \(\Delta t\), the falling-edge response has sufficient time to reset the pixel state, enabling a more stable response to subsequent pulses.
    }
    \label{fig:short_pulse_recovery}
\end{figure*}

Figure~\ref{fig:short_pulse_recovery} shows the evolution of the event stream as a function of pulse separation. 
For sufficiently large values of \(\Delta t\) (see Figure \ref{fig:short_pulse_recovery} (c)), each optical pulse generates a stable sequence of positive and negative events, indicating that the internal pixel state is fully restored before the arrival of the next pulse. As the pulse separation is reduced, the negative-polarity response progressively weakens, while the positive response initially remains largely unaffected, resulting in an increasing polarity imbalance. For the shortest pulse separations, Figure \ref{fig:short_pulse_recovery} (a), both the event rate and the number of generated negative events decrease markedly, demonstrating that the sensor is no longer able to recover completely between successive optical transitions.

This behavior is consistent with the polarity-dependent reset dynamics discussed in the frequency characterization. Since each positive event updates the internal reference level to a higher illumination value, incomplete generation of negative events prevents the reference from returning to its initial state before the next excitation. Consequently, successive pulses no longer produce independent bursts of events, and the sensor progressively loses sensitivity to repeated optical transitions.

\subsection{Pulse Reconstruction Fidelity}
\label{subsec:pulse_reconstruction}
The final pulse-response experiment was designed to evaluate how accurately the temporal structure of an optical pulse can be reconstructed from the event stream. In this case, pulses were again separated by \(20\,\mathrm{ms}\), ensuring that consecutive responses did not overlap, while the pulse width itself was varied. For each acquisition, the event stream was converted into a signed event-count signal sampled with \SI{1}{\micro\second} temporal resolution,
\begin{equation}
    s_m = N_{\mathrm{POS}}(m) - N_{\mathrm{NEG}}(m),
\end{equation}
where \(N_{\mathrm{POS}}(m)\) and \(N_{\mathrm{NEG}}(m)\) are the numbers of positive and negative events detected in the temporal bin \(m\). A cumulative reconstruction signal was then obtained as \(q_m = \sum_{j \leq m} s_j\). The reconstructed pulse was compared with the imposed optical waveform by evaluating the timing of the reconstructed rising and falling transitions and the resulting pulse width. This experiment also extends the sinusoidal reconstruction performed in the previous section to a more abrupt temporal stimulus. In the sinusoidal case, the illumination varies continuously, and the generated events are distributed over the full period of the waveform. In contrast, a pulse concentrates the optical change into a rising edge and a falling edge, so the reconstructed waveform depends more strongly on the timing and balance of the positive and negative responses.

To evaluate this, the LED was driven with isolated voltage pulses with widths from \SI{10}{\micro\second} to \SI{1000}{\micro\second}. For each pulse width, the signed events were cumulatively summed across repeated acquisitions to obtain the mean reconstructed response. The recovered pulse duration was then estimated from the decay of the reconstructed signal after the peak, using the final value within the displayed time window as the reference level. This allowed the apparent pulse width measured from the EVS reconstruction to be compared directly with the pulse width applied to the LED.

\begin{figure*}[h!]
    \centering

    \begin{subfigure}{\linewidth}
        \centering
        \includegraphics[width=0.9\linewidth]{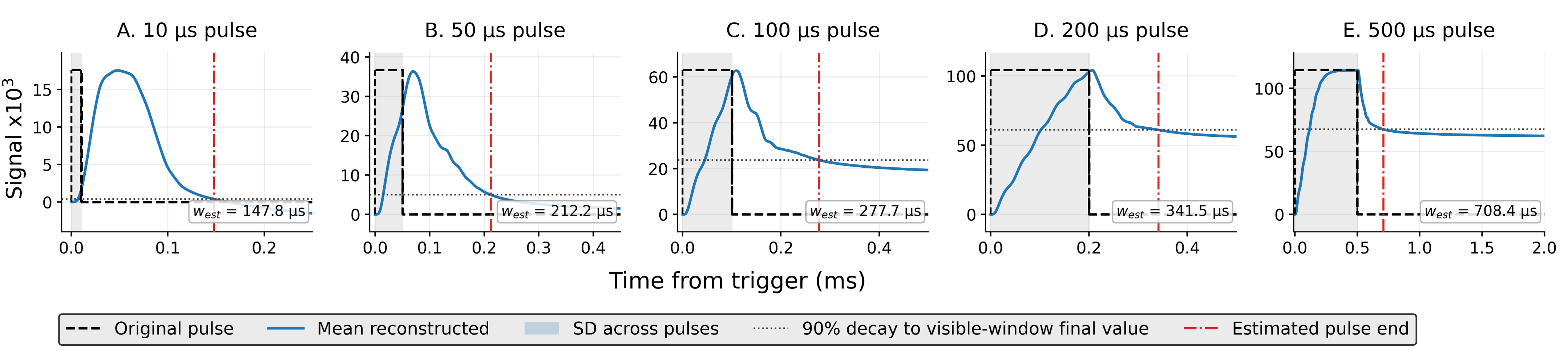}
        \caption{}
        \label{fig:pulse_recon_traces}
    \end{subfigure}

    \vspace{0.6em}

    \begin{subfigure}{\linewidth}
        \centering
        \includegraphics[width=0.6\linewidth]{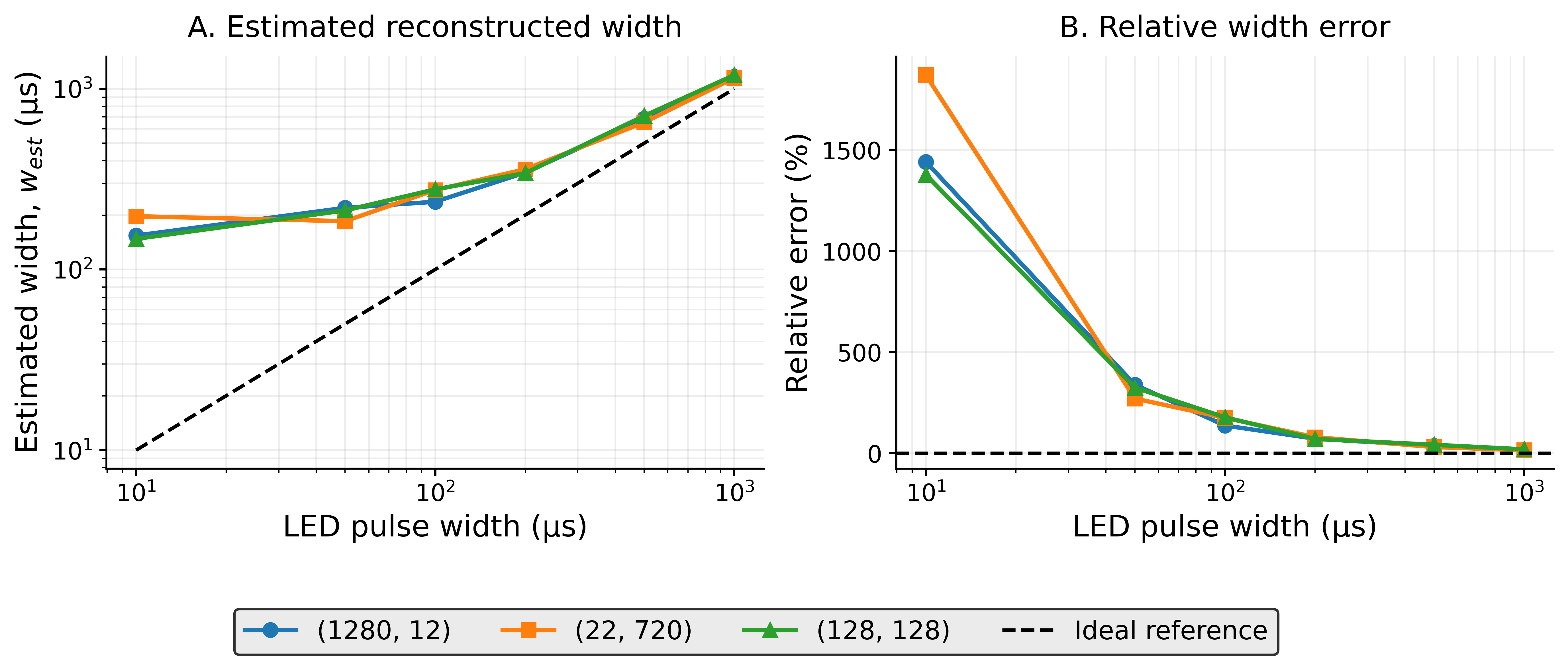}
        \caption{}
        \label{fig:pulse_recon_width_metrics}
    \end{subfigure}

    \caption{Pulse reconstruction and pulse-width estimation from signed EVS event accumulation: (a)LED Pulse-width reconstruction via event accumulation. Mean reconstructed responses were obtained by cumulatively summing signed EVS events across repeated LED pulse acquisitions for the 128 $\times$ 128 pixels ROI. Columns show LED command pulse widths from 10 to \SI{1000}{\micro\second}. Dashed black traces indicate the nominal LED input pulse (the ideal reference pulse), scaled to the maximum reconstructed response in each panel. Blue traces show the mean reconstructed response, and the shaded region shows the standard deviation across repeated pulses. The dotted horizontal line marks the 90\% decay threshold relative to the peak and the final value within the displayed time window, and the red dash-dotted vertical line indicates the estimated reconstructed pulse end. The estimated reconstructed width is measured from the trigger onset to this threshold-crossing time; (b) Pulse-width estimation metrics extracted from the reconstructed responses. The estimated reconstructed width, $w_{\mathrm{est}}$, is compared against the commanded LED pulse width, $w_{\mathrm{LED}}$, for each region of interest. The ideal response is shown by the dashed reference line in the width-comparison plot and by the zero-error line in the relative-error plot. Relative error was computed as $100 \times (w_{\mathrm{est}} - w_{\mathrm{LED}}) / w_{\mathrm{LED}}$.}
    \label{fig:pulse_recon}
\end{figure*}

Figure \ref{fig:pulse_completion} (a) presents representative reconstructed waveforms together with the measured pulse widths. For long optical pulses, the reconstructed signals accurately reproduce the imposed excitation, exhibiting well-defined rising and falling transitions and pulse durations close to the input values. In this regime, the intrinsic temporal broadening of the EVS response represents a small fraction of the total pulse duration, so the accumulated signed events provide a more faithful representation of the optical waveform. However, since the sensor produces more positive than negative events, the reconstructed signal doesn't necessarily approach 0 after the falling edge. 

As the pulse width decreases, however, the reconstructed transitions become progressively broader and the measured pulse duration increasingly exceeds the true optical pulse width. For LED pulse widths of \SI{10}{\micro\second} and \SI{50}{\micro\second}, the temporal spreading associated with the positive and negative responses dominates the reconstruction. The falling-edge response is particularly important, since negative events continue to be generated after the illumination returns to the lower state. Consequently, the reconstructed waveform contains a long decay tail that is not present in the original optical signal.

The influence of ROI geometry follows the same trend observed throughout this work, as is further described in the Supplementary Materials. ROIs spanning more sensor rows exhibit greater temporal spreading due to increased contributions from row arbitration and event serialization, resulting in broader reconstructed transitions. In contrast, more compact ROIs preserve sharper transitions and therefore provide more accurate pulse-width estimates, particularly for short optical pulses.

These results are consistent with the short-pulse recovery measurements discussed previously. Even when sufficient time is provided between consecutive pulses to avoid progressive degradation of the response, the temporal structure reconstructed from the EVS event stream remains broadened relative to the original optical excitation. Therefore, although the sensor can detect illumination transitions on microsecond timescales, accurate reconstruction of microsecond optical waveforms is limited by the temporal spreading of the positive and negative event responses.

\section{Discussion}
To synthesize the present work, Table \ref{tab:summary_metrics} summarizes the main sensor response characteristics identified throughout the experimental characterization. Rather than focusing exclusively on absolute quantitative limits, this summary highlights the practical mechanisms and guidelines to consider for improved performance when designing scientific measurement systems based on event-based vision sensors. 

Indeed, we note that the effective temporal performance of these devices depends strongly on the specific experimental configuration, including the background illumination level, optical contrast, modulation frequency, event density, ROI geometry, and the required trade-off between spatial coverage and temporal fidelity. Consequently, metrics such as response duration, reconstruction accuracy, or usable modulation bandwidth cannot be interpreted as fixed intrinsic sensor properties independent of operating conditions. For this reason, the present work emphasizes the physical origins and experimental consequences of the observed behaviors rather than defining application-independent performance limits. In scientific instrumentation applications, the relevant operating regime is highly dependent on the characteristics of the measured phenomenon and the constraints imposed by the acquisition system. 

\begin{table*}[t]
\centering
\caption{Summary of the main temporal-response characteristics observed in the EVS characterization experiments.}
\vspace{2mm}
\label{tab:summary_metrics}
\renewcommand{\arraystretch}{1.25}
\arrayrulecolor{black}

\begin{tabular}{|>{\columncolor[HTML]{FFEDD8}\centering\arraybackslash}m{4.5cm}|>{\centering\arraybackslash}m{12.5cm}|}
\hline
\rowcolor[HTML]{FFCE93}
\textbf{Parameter} & \textbf{Key observations} \\ \hline

First-event latency (\(T_{\mathrm{first}}^{(p)}\)) &
Below \SI{5}{\micro\second} for the first detected event, with weak dependence on ROI geometry. \\ \hline

ROI response completion time (\(T_{\mathrm{first,all}}^{(p)}\))&
Strongly dependent on ROI geometry and readout serialization, corresponding to the time required for all pixels in the ROI to generate a first response. \\ \hline

Full event-stream duration (\(T_{\mathrm{all}}^{(p)}\)) &
Pixels may generate multiple events for a single illumination transition, producing temporally extended event bursts that define the effective response width. \\ \hline

Temporal edge detection fidelity &
Optical pulse transitions remain detectable for pulse widths down to approximately \SI{50}{\micro\second}. \\ \hline

Waveform reconstruction fidelity &
Accurate pulse-shape reconstruction is only achieved for illumination pulses longer than approximately \SI{200}{\micro\second}. \\ \hline

Pulse-train recovery &
Closely spaced pulses can prevent full OFF-event generation and pixel reset, causing progressive degradation of the reconstructed response, leading to pixel response stagnation. \\ \hline

Background illumination level &
Higher background illumination generally increases event activity and improves response robustness under high-contrast conditions. \\ \hline

Illumination contrast variation &
Lower illumination contrast reduces event generation and limits the effective temporal bandwidth. \\ \hline

\end{tabular}
\end{table*}

\section{Conclusion}

This work presented a systematic characterization of an IMX636-based event camera under controlled sinusoidal and pulsed optical excitation to evaluate its suitability as a high-speed optical measurement instrument. The results show that event cameras can provide microsecond-scale sensitivity to abrupt illumination changes, with first-event latencies below \SI{5}{\micro\second} in the tested conditions. However, this earliest detectable response does not represent the complete temporal response of the sensor. The event stream generated by an optical transition is shaped by photoreceptor settling, contrast-threshold crossings, refractory behavior, polarity-dependent reset dynamics, and ROI-dependent readout serialization.

For sinusoidal illumination, the sensor recovered periodic optical signals over a broad range of modulation frequencies when the illumination level and modulation contrast were favorable. The lock-in amplitude indicates that a coherent response at the commanded modulation frequency remains measurable up to approximately \(80~\mathrm{kHz}\), and weak spectral components can still be identified at higher frequencies. However, the significance of this component relative to the surrounding spectral background decreases as the modulation frequency increases. This is reflected in the reduction of local SNR, the progressive loss of waveform contrast in the reconstructed time-domain signal, and the increasing imbalance between positive and negative events. Therefore, the presence of a spectral peak at the imposed frequency should be interpreted as evidence of frequency detectability rather than of faithful waveform reconstruction. The effective frequency response is consequently not governed by a single bandwidth value, but depends on the required criterion, namely, whether the goal is to detect the modulation frequency or to reconstruct the optical waveform with high fidelity.

The pulse measurements further demonstrated the distinction between detecting an optical transition and faithfully reconstructing the corresponding optical waveform. Isolated optical edges generated event bursts that extended well beyond the physical LED transition time, while short pulse trains revealed that incomplete OFF-event generation can prevent full pixel reset and lead to progressive response degradation. In the variable-width pulse experiment, reconstructed pulse durations were significantly broadened for short optical pulses and approached the commanded LED widths only when the pulse duration became sufficiently long relative to the sensor response and readout times. Thus, microsecond-scale event detection is clearly possible but should not be interpreted as microsecond-scale waveform fidelity.

The dependence on ROI geometry also highlights the importance of matching the acquisition configuration to the measurement task. When spatially extended illumination activates many pixels simultaneously, the readout architecture becomes part of the temporal response. Wider or taller (multiple-row) ROIs can yield different event densities and serialization patterns, which affect response duration, waveform smoothness, and reconstruction accuracy. Therefore, it is clear that ROI selection is also a temporal design parameter in event-based optical measurements that must be taken into consideration.

Overall, these results reinforce the promise of event-based sensors for high-speed optical instrumentation, particularly when the goal is to detect rapid transitions or recover periodic signals under suitable operating conditions. However, quantitative waveform reconstruction requires careful interpretation of the event stream and cannot rely solely on timestamp precision. The characterization presented here provides practical guidelines for selecting illumination levels, modulation contrasts, ROI geometries, and pulse spacings, and emphasizes the need to distinguish between first-event latency, full response duration, recoverable frequency content, polarity stability, and waveform reconstruction fidelity when designing scientific experiments based on event-based vision sensors.

\section{Acknowledgments}
Tomás Lopes and Joana Teixeira acknowledge the support of the Foundation for Science and Technology (FCT), Portugal, through Grants 2024.01830.BD and 2024.00426.BD, respectively. Nuno A. Silva acknowledges the support of FCT under the grant \\ 2022.08078.CEECIND/CP1740/CT0002
(https://doi.org/10.54499/2022.08078.CEECIND/CP1740/CT0002). Tiago D. Ferreira acknowledges the support of FCT under the grant 2024.10684.CEECIND. This work was co-financed by National Funds through the FCT - Fundação para a Ciência e a Tecnologia, I.P. (Portuguese Foundation for Science and Technology) within the project QuAC, with reference 2024.16086.PEX (https://doi.org/10.54499/2024.16086.PEX).

\bibliographystyle{IEEEtran}
\bibliography{sample}

\section{Biography Section}

\begin{IEEEbiographynophoto}{Tomás Lopes}
is currently a PhD student at the University of Porto, in collaboration with INESC TEC, Portugal. His current research interests include optical computing and sensing and spectral imaging. 
\end{IEEEbiographynophoto}

\begin{IEEEbiographynophoto}{Joana M. Teixeira}
is currently a PhD student at the University of Porto, in collaboration with INESC TEC, Portugal. Her current research interests include optical sensing, spectral imaging, and microfluidics. 
\end{IEEEbiographynophoto}

\begin{IEEEbiographynophoto}{Tiago D. Ferreira}
received his PhD degree of Physics in 2024 from the University of Porto, Portugal. Now, he is an assistant researcher at the Centre for Applied Photonics, INESC TEC. His current research interests include paraxial fluids of light, quantum imaging and sensing, and optical computing and sensing. 
\end{IEEEbiographynophoto}

\begin{IEEEbiographynophoto}{Catarina S. Monteiro}
Received her PhD degree of Engineering Physics from the University of Porto, Portugal in 2023. Now, she is an assistant researcher at CAP, INESCTEC, Portugal, and assistant professor at the Department of Physics and Astronomy of the Faculty of Sciences of the University of Porto. Her current research interests include optical sensing, quantum sensing, and spectral imaging. 
\end{IEEEbiographynophoto}

\begin{IEEEbiographynophoto}{Pedro A.S. Jorge}
completed his PhD degree in Physics at the University of Porto in 2006, in collaboration with the Department of Physics and Optical Sciences at the University of North Carolina at Charlotte, USA. He is currently Senior Researcher and team leader at the Centre for Applied Photonics of INESC TEC, where he leads research on photonic sensing, optical instrumentation, spectroscopy and spectral imaging. He is also Assistant Professor at the Department of Physics and Astronomy of the Faculty of Sciences of the University of Porto and, since 2021, Director of the MSc in Physics Engineering. 
\end{IEEEbiographynophoto}

\begin{IEEEbiographynophoto}{Nuno A. Silva}
completed his PhD degree in Physics at the University of Porto in 2019, in collaboration with INESC TEC. Now he is Senior Researcher at the Centre for Applied Photonics of INESC TEC, doing research on paraxial fluids of light, quantum imaging and sensing, optical computing and sensing, spectroscopy and spectral imaging. He is also Assistant Professor at the Department of Physics and Astronomy of the Faculty of Sciences of the University of Porto and, since 2026. 
    
\end{IEEEbiographynophoto}

\vfill

\end{document}